\documentclass[conference,10pt,a4paper]{IEEEtran}
\pdfoutput=1 
\usepackage[T1]{fontenc}
\usepackage[utf8]{inputenc}
\usepackage[british]{babel}
\usepackage{amsmath}
\usepackage{textcomp}
\usepackage{amssymb}
\usepackage{url}
\usepackage{hyperref}
\usepackage{cleveref}
\usepackage{todonotes}
\usepackage{booktabs}
\usepackage[noprefix]{nomencl}	
\usepackage{stmaryrd}

\graphicspath{{figures/}}

\usepackage[activate={true,nocompatibility},final,kerning=true,tracking=true,spacing=true]{microtype}
\microtypecontext{spacing=nonfrench}

\newcommand{\x}{$\times$}
\newcommand{\eg}{e.g.\ }
\newcommand{\ie}{i.e.\ }
\newcommand{\etal}{et al.\ }
\renewcommand{\v}[1]{\ensuremath{\mathbf{#1}}} 
\newcommand{\gv}[1]{\ensuremath{\mbox{\boldmath$ #1 $}}}  
\newcommand{\grad}[1]{\gv{\nabla} #1} 
\renewcommand{\div}[1]{\gv{\nabla} \cdot #1} 
\renewcommand{\d}[2]{\frac{d #1}{d #2}} 
\newcommand{\pd}[2]{\frac{\partial #1}{\partial #2}} 
\newcommand{\nomunit}[1]{\renewcommand{\nomentryend}{\hspace*{\fill}#1}}

\crefname{chapter}{Chapter}{Chapters}
\crefname{section}{Section}{Sections}
\crefname{appendix}{Appendix}{Appendices}
\crefname{subsection}{Section}{Sections}
\crefname{subsubsection}{Section}{Sections}
\crefname{equation}{}{}
\crefname{figure}{Fig.}{Figs.}
\crefname{table}{Table}{Tables}
\crefname{subfigure}{Fig.}{Figs.}
\crefname{listing}{Listing}{Listings}

\makenomenclature

\begin{document}
\title{Experimental and computational studies of water drops falling through
  model oil with surfactant and subjected to an electric field\vspace{-20pt}}
\author{\IEEEauthorblockN{{\AA}smund Ervik\IEEEauthorrefmark{1}, Svein
    Magne Helles{\o}\IEEEauthorrefmark{2}, Svend Tollak
    Munkejord\IEEEauthorrefmark{2} and Bernhard
    M\"{u}ller\IEEEauthorrefmark{1}}
\IEEEauthorblockA{\IEEEauthorrefmark{1} Dept. of Energy and Process
  Engineering, Norwegian University of Science and Technology, NO-7491
  Trondheim, Norway\\ Email: asmund.ervik@ntnu.no}
\IEEEauthorblockA{\IEEEauthorrefmark{2} SINTEF Energy Research, P.O. Box
  4761 Sluppen, NO-7465 Trondheim, Norway}
\vspace{10pt}
\IEEEauthorblockA{
  {\large This paper is to be presented at the} \\ {\large 2014 IEEE 18th International Conference
  on Dielectric Liquids (ICDL).} \\ {\large As such, this paper is Copyright
    \textcopyright~2014 IEEE.}
}

}

\maketitle
\begin{abstract}
The behaviour of a single sub-millimetre-size water drop falling through a
viscous oil while subjected to an electric field is of fundamental importance
to industrial applications such as crude oil electrocoalescers. Detailed
studies, both experimental and computational, have been performed
previously, but an often challenging issue has been the characterization of the
fluids. As numerous authors have noted, it is very difficult to have a
perfectly clean water-oil system even for very pure model oils, and the
presence of trace chemicals may significantly alter the interface behaviour. 
In this work, we consider a
well-characterized water-oil system where controlled amounts of a surface
active agent (Span 80) have been added to the oil. This addition dominates any
trace contaminants in the oil, such that the interface behaviour can also be
well-characterized. We present the results of experiments and corresponding
two-phase-flow simulations of a falling water drop covered in surfactant and
subjected to a monopolar square voltage pulse. The results are compared and
good agreement is found for surfactant concentrations below the critical micelle
concentration.
\end{abstract}
\printnomenclature[1.5cm]

\hfill \today

\section{Introduction}
The study of a single fluid drop falling through some other fluid has a long
tradition of study; the physics of raindrops has been debated ever since
Aristotle's ideas on the subject in ancient times. Pioneers in this field
include Rayleigh, Kelvin, Stokes, Reynolds and Worthington, the 
latter of whom dedicated most of his career to the study of splashing drops and
pioneered the use of flash photography for this purpose.

In spite of such prolonged study, some details elude our understanding still
today. This basic phenomenon is also relevant for various industrial
applications. In this paper our focus is on sub-millimetre-size water drops
falling through a more viscous medium of slightly lower density, i.e. oil, with
an eye towards understanding the processes that govern the operation of
electrocoalescers used for crude oil processing \cite{eow2001}.

In this context, a particular hurdle is the characterization of the
two fluids and whether or not the drop surface is clean. To
overcome this uncertainty we consider a system where a surface-active agent
(Span 80) is added such that it dominates existing impurities. The whole
system is then well-characterized in terms of viscosities, densities and
interfacial tension as a function of surfactant concentration. We report the
outcome of experiments with a falling drop in this fluid system, where in
addition a monopolar square voltage pulse is applied to deform the drop. These
experiments are compared with two-phase flow simulations where the electric
field and the surface-active agent are taken into account.

\section{Theory}
The equations that govern the two-phase flow system under consideration are the
incompressible Navier-Stokes equations:
\begin{align}
  \div\v{u} &= 0 \label{eq:ns-divfree} \\
	\pd{\v{u}}{t}+(\v{u}\cdot\grad)\v{u} &= - \frac{\grad{p}}{\rho}
  + \frac{\eta}{\rho}\grad^2\v{u} +
	\v{f}
	\label{eq:ns}
\end{align}%
The above equations hold for a single fluid, or separately in each fluid for an
immiscible two-phase system.  Across the interface between the fluids there will
be discontinuities in some quantities, \eg the pressure. These discontinuities
are enforced in a sharp manner using the ghost fluid method and the level-set
method in the simulations.
We will refer to the drop fluid as having properties with subscript $_1$ and
the ambient (bulk) fluid as having properties with subscript $_2$.
\nomenclature{$\v{u}(\v{x})$}{Velocity field of a fluid.\nomunit{m/s}}%
\nomenclature{$\eta$}{Dynamic viscosity of a fluid.\nomunit{Pa$\cdot$s}}%
\nomenclature{$\rho$}{Density of a fluid.\nomunit{kg/m$^{3}$}}%
\nomenclature{$p(\v{x})$}{Pressure of a fluid.\nomunit{Pa}}%
\nomenclature{$\v{f}$}{External acceleration.\nomunit{m/s$^2$}}%

For the fluids and drop sizes we consider, the Reynolds number will be
$\rm{Re} \lesssim 1$ such that the flow is laminar. The terminal velocity of
a single falling clean drop can then be predicted by the Hadamard-Rybczynski
formula \cite{hadamard1911,rybzynski1911}
\begin{equation}
  \v{v}_{\rm{T,HR}} = \frac{2(\rho_1 - \rho_2) \v{g} R^2(\eta_1 + \eta_2)}{3\eta_2(3\eta_1 + 2\eta_2)}
  \label{eq:hadamard}
\end{equation}
while the terminal velocity of a contaminated drop will be closer to that of
a falling rigid sphere, predicted by the Stokes formula \cite{lamb1945}
\begin{equation}
  \v{v}_{\rm{T,S}} = \frac{2(\rho_1 - \rho_2) \v{g} R^2}{9\eta_2} \rm{.}
  \label{eq:stokes}
\end{equation}
Note that $\v{g}$ is a vector and the resulting velocity vector will have the
correct sign for both a falling and rising drop. We note the curious 
fact that for a given $R$ these velocities can never be
equal, as long as the viscosities are both non-zero and finite. This allows an
easy way of distinguishing a clean system from a contaminated one. In practice,
it is easier to consider a log-log plot of the drag coefficient versus the drop
Reynolds number, since both the Stokes theory and the Hadamard-Rybczynski theory
predict these values should fall on a straight line, but they predict two
different (parallel) lines. 

In addition to the fluid flow, an electric field is imposed on the system. In
the experiments this is done by applying a monopolar square voltage pulse to
plane electrodes at the top and bottom of the fluid chamber. The electric
potential $\Psi$ can be calculated from boundary conditions corresponding to the
voltage on the two electrodes and from the permittivity $\epsilon\epsilon_0$ in the two
fluids by a Laplace equation
\begin{equation}
  \div(\epsilon\epsilon_0\grad\Psi) = 0 \rm{.}
  \label{eq:laplace}
\end{equation}
This is the simplest model for a conducting drop in a dielectric medium: we
assume that both media are perfect dielectrics but with a high permittivity
ratio \cite{melcher1969}. 
\nomenclature{$R$}{Radius of a drop.\nomunit{m}}%
\nomenclature{$\v{g}$}{Gravitational acceleration.\nomunit{m/s$^2$}}%
\nomenclature{$\sigma$}{Electrical conductivity.\nomunit{S/m}}%
\nomenclature{$\Psi(\v{x})$}{Electric potential.\nomunit{V}}%
\nomenclature{$\epsilon$}{Relative permittivity.\nomunit{-}}%
\nomenclature{$\epsilon_0$}{Vacuum permittivity.\nomunit{F/m}}%

\section{Methods}
\subsection{Experimental methods}
The experiments were conducted with brine water drops falling in Marcol 52
oil. The brine water was prepeared by adding 3.5~\% (by weight) NaCl to
highly purified water.  Marcol 52 is a purified mixture of liquid saturated
hydrocarbon that has a very low content of surface-active components.
Different amounts of the surfactant Span 80 was added to the oil, completly
dominating as a surfactant.

The density of water and oil were measured on a DMA 5000 densitymeter from
Anton Paar. The vicsocity of the oil was measured using a Anton Paar MCR 102
rheometer. The viscosity of the water was assumed to be identical to
established values for sea water. The experiments were conducted at
21.5\textdegree C. At this temperature the viscosity and density of the water
were 1.03 mPa$\cdot$s and 1023.6 kg/m$^{3}$, respectively, and those of the
oil were 12.4 mPa$\cdot$s and 832.3 kg/m$^{3}$, respectively.

The interfacial tension $\gamma$ between water and oil was measured
using a SIGMA 703D tensiometer with a DuNuoy ring. The results for
the different amounts of surfactant are given in \cref{tab:IFT}. The
concentration is given in percent by weight, denoted by wt\%.
 \begin{table}[!t]
  \caption{IFT between water and oil for different surfactant concentrations}
  \label{tab:IFT}
  \centering
\begin{tabular}{cc}
  \toprule
Span 80 concentr. [wt\%] & Interfacial tension [mN/m] \\ 
\midrule
0.030 & 10.0 \\ 
0.020 & 10.1 \\ 
 0.015 & 13.9 \\ 
0.010 & 18.8 \\ 
0.001 & 29.4 \\ 
\bottomrule
\end{tabular} 
\end{table}
These results indicate that the critical micelle concentration for our system
is 0.020 wt\%. We will restrict ourselves to concentrations below this. 

\begin{figure}[!t] 
   \centering
   \includegraphics[width=0.5\linewidth]{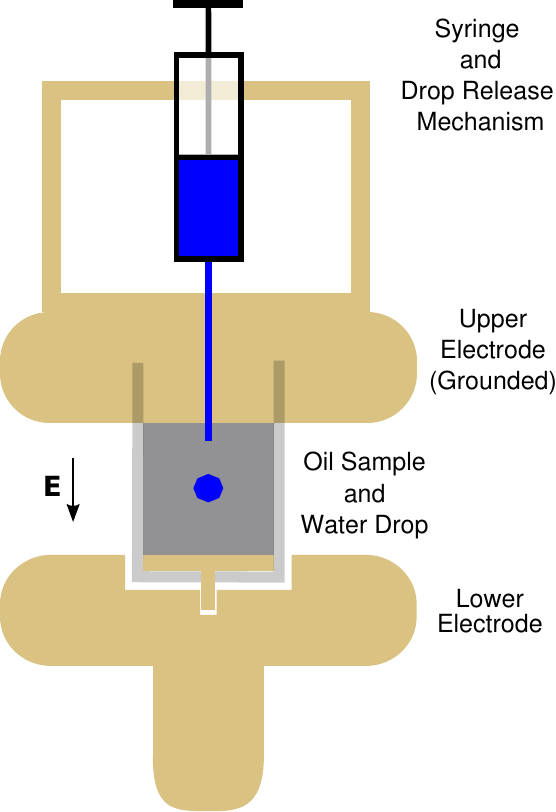}
   \caption{Side view of the experimental setup. The electric field is aligned
   with the gravitational field.}
   \label{fig:test-cell}
\end{figure}

The experiment consisted of letting water drops fall in the 15 mm gap between
an upper and a lower horizontal electrode. A monopolar square voltage was
applied to the lower electrode, creating an electrical field $\v{E}$ that
distorted the drop. The voltage was generated using a Stanford Research DS340
signal generator connected to a TREK 2020B high voltage amplifier. Field
strengths were varied between 266.7 V/mm and 666.7 V/mm for different cases,
as given in the following. A side view of the experimental setup is shown in 
\cref{fig:test-cell}.
 
\begin{figure}[!t] 
    \centering
    \includegraphics[width=0.5\linewidth]{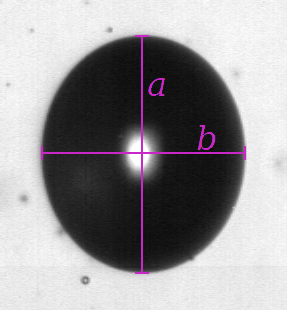}
    \caption{Example of image captured of elongated drop with major and minor
      axes $a$ and $b$ superimposed.}
    \label{fig:frame_180}
  \end{figure}

A high-speed Cheetah CL near infrared camera with a Infinity KS2 long-range
microscope lens was used to record movies of the drops as they fell and
were deformed by the field. The camera had a resolution of $640\times 512$ pixels and could
record 1730 frames per second. The position and deformation of the drops
were extracted from the movies using the Spotlight image analysis software
\cite{spotlight}. In particular, the vertical velocity and the ratio of the
major and minor axis $a/b$ were determined, as shown in
\cref{fig:frame_180} where a slightly elongated water drop is seen with the axes
superimposed.
 
The experimental procedure consisted of these steps:\\
\textbf{1.} Create
a water drop of the desired size at the tip of the glass needle using the 
screw-in plunger. The field of view of the camera is at the needle tip. \\
\textbf{2.} Once a
drop is created, move the test cell such that the field of view of the
camera is at the center of the electrode gap. \\
\textbf{3.} Arm the camera to start recording on trigger.  \\
\textbf{4.} Release the drop from the tip of the glass needle. \\
\textbf{5.} Once the drop comes into view of the camera, trigger the camera and
voltage source. \\ 
\textbf{6.} The camera records a movie of the drop. \\
\textbf{7.} Extract the drop position and shape from the recorded movie.

\subsection{Simulation and numerical methods}
The Navier-Stokes equations (\ref{eq:ns-divfree}) and (\ref{eq:ns}) are solved
numerically using an in-house code. All reported simulations are done using 2D
axisymmetry. The equations are discretized with a finite-difference method,
specifically WENO \cite{liu1994} for the convective terms and central
differences for the viscous terms. The time integration is done with an explicit
Runge-Kutta method (SSPRK (2,2) in the terminology of Gottlieb \etal
\cite{gottlieb2009}). A standard projection method \cite{chorin1968} is used to
determine the pressure field such the velocity field is solenoidal, giving
a Poisson equation for the pressure $p$. The solution of this Poisson equation
demands the majority of the computing time; it is solved here using the
Bi-Conjugate Gradient Stabilized method \cite{vandervorst1992} with the
BoomerAMG preconditioner \cite{henson2000}. The same methods are used to solve
the Laplace equation for the electric field. We employ the PETSc and Hypre
libraries for these methods \cite{balay1997}, \cite{hypre}.

As stated, the Navier-Stokes equations are valid in each phase separately. The
interface between the phases is tracked using the level-set method, see
\cite{ervik2014} and references therein for a thorough description of how we
employ this method. To couple the two phases, the ghost-fluid method (GFM)
\cite{GFMNavier,teigen2010} is used in the simulations to enforce the jumps in 
different quantities (like pressure) across the interface:
\begin{align}
\llbracket \v{u} \rrbracket &=0 \rm{,}\\
\label{eq:pjump}
\llbracket p \rrbracket &=2\llbracket \eta \rrbracket \v{n}\cdot \grad \v{u}\cdot
\v{n} + \v{n}\cdot \llbracket \v{M} \rrbracket \cdot \v{n} + \gamma\kappa \rm{,}\\
\llbracket \grad p \rrbracket &=0 \rm{,} \\
\llbracket \Psi \rrbracket  &= 0 \rm{,}\\
\llbracket \epsilon \v{n}\cdot \grad \Psi \rrbracket &=0 \rm{.}\\
\label{eq:gradujump}
\llbracket \eta \grad \v{u}\rrbracket &=\llbracket \eta \rrbracket \Bigl(
(\v{n}\cdot \grad \v{u}\cdot \v{n})\v{n}\v{n}+(\v{n}\cdot \grad \v{u}\cdot \v{t})\v{n}\v{t} \nonumber \\
&\qquad\;+ (\v{n}\cdot \grad \v{u}\cdot \v{t})\v{t}\v{n}+(\v{t}\cdot \grad \v{u}\cdot \v{t})\v{t}\v{t} \Bigr) \\
&\quad + (t\cdot \llbracket \v{M}\rrbracket \cdot \v{n}) + \v{t}\v{n} (\v{t}\cdot
\grad_\iota \gamma)\v{t}\v{n} \rm{,} \nonumber
\end{align}
Here $\iota$ denotes the interface between the two fluids (so $\grad_\iota$ is the
gradient along the interface) and $\v{M}$ is the
Maxwell stress tensor calculated from the electric field
$\v{E}$. This is again calculated from the potential $\Psi$ which we obtain by
solving \cref{eq:laplace} numerically. With the same approximation as in
\cref{eq:laplace}, the Maxwell stress tensor is  
\begin{equation}
  \v{M} = \epsilon\epsilon_0\Bigl(\v{E}\v{E}-\frac{1}{2}(\v{E}\cdot\v{E})\v{I}\Bigr) \rm{,}
  \label{eq:Mtensor}
\end{equation}
and $\v{n}$ and $\v{t}$ are the normal and tangent vectors to the
interface, and formulas like
$\v{n}\v{n}$ and $\grad\v{u}$ denote tensors formed by the outer product. Our
convention is that the normal vector points out from a drop and that the
jump $\llbracket\cdot\rrbracket$ is the difference going from the outside to the
inside, \eg $\llbracket \eta \rrbracket = \eta_2 - \eta_1$. 

This formulation takes into account the pressure difference due to
interfacial tension $\gamma$ in \cref{eq:pjump}, the applied
electric field, and the Marangoni effect that arises from an interfacial
tension gradient along the interface in \cref{eq:gradujump}. The surfactant
concentration $\xi$ determines $\gamma$ through the Langmuir equation
\begin{equation}
  \gamma(\xi)=\gamma_0\left(1+\beta\ln(1-\frac{\xi}{\xi_\infty}) \right) \rm{;}
  \label{eq:surfactant}
\end{equation}
where $\beta =R T \xi_\infty/\gamma_0$ is the interfacial elasticity and
$\xi_\infty$ is the maximum possible concentration of surfactant.  We consider
here an insoluble surfactant which has zero concentration away from the
interface. Furthermore, we restrict ourselves to surfactant concentrations 
which are below the critical micelle concentration (0.02 wt\% for our
system). An insoluble surfactant
is then a good approximation when both the bulk and the interface
Peclet numbers, $\rm{Pe}_{2} = \v{v}_TR/D_{\xi,2}$ and $\rm{Pe}_{\iota}
= \v{v}_TR/D_{\xi,\iota}$, are high \cite{sadhal1983}. Here
$\v{v}_T$ is the terminal velocity and $D_{\xi,2},D_{\xi,\iota}$ are
diffusion constants for the surfactant in the bulk phase and on the interface,
respectively. 
There is only sparse literature data on these diffusion constants, but they seem
to be around $10^{-10}$ m/s$^2$ \cite{wang1997msc}. We will assume here that the
two diffusion constants (and thus the Peclet numbers) are equal, writing simply
$D_{\xi}$. Tests revealed that our simulations are not very sensitive to the
value, with values of $10^{-10}$ and $5\cdot10^{-7}$ giving indistinguishable
results, so a value of $10^{-10}$ was subsequently used. We may understand this
low sensitivity when we compute the respective Peclet numbers, they are 250 and
50 000 for these values of the diffusion constant, which are both $\gg 1$. This
indicates that an insoluble surfactant should be a good approximation.

The surfactant is transported along the interface according to
an advection-diffusion equation which can be expressed in
the same coordinates as all the other equations, by using the curvature of the
interface $\kappa$ and the normal vector $\v{n}$,
\begin{align}
  \d{\xi}{t} +& \v{u} \cdot \grad \xi - \v{n}\cdot \grad \v{u} \cdot \v{n}\, \xi = \nonumber \\ 
  &D_{\xi} \Bigl( \grad^2\xi - \v{n} \cdot\grad\grad\cdot\v{n}\,\xi
  + \kappa(\v{n}\cdot\grad \xi)\Bigr)\rm{,}
\end{align}
where the first two terms on the left-hand
side and the first term on the right-hand side are the ordinary terms in an
advection-diffusion equation, and the remaining terms constitute the restriction
of this equation to the interface. 
\nomenclature{$\gamma$}{Interfacial tension.\nomunit{N/m}}%
\nomenclature{$\xi$}{Surfactant concentration.\nomunit{wt\% (\% by weight)}}%

The surfactant transport equation, as well as the level-set equations for
interface tracking, are discretized with finite-difference methods and
integrated in time using explicit Runge-Kutta methods. We refer to Teigen
\etal \cite{teigen2010,teigen2010a,ICDL2011} for a thorough discussion and
validation of these models.

\section{Results}
\subsection{Terminal velocity}
\begin{figure}[!t]
  \centering
  \includegraphics[width=\linewidth]{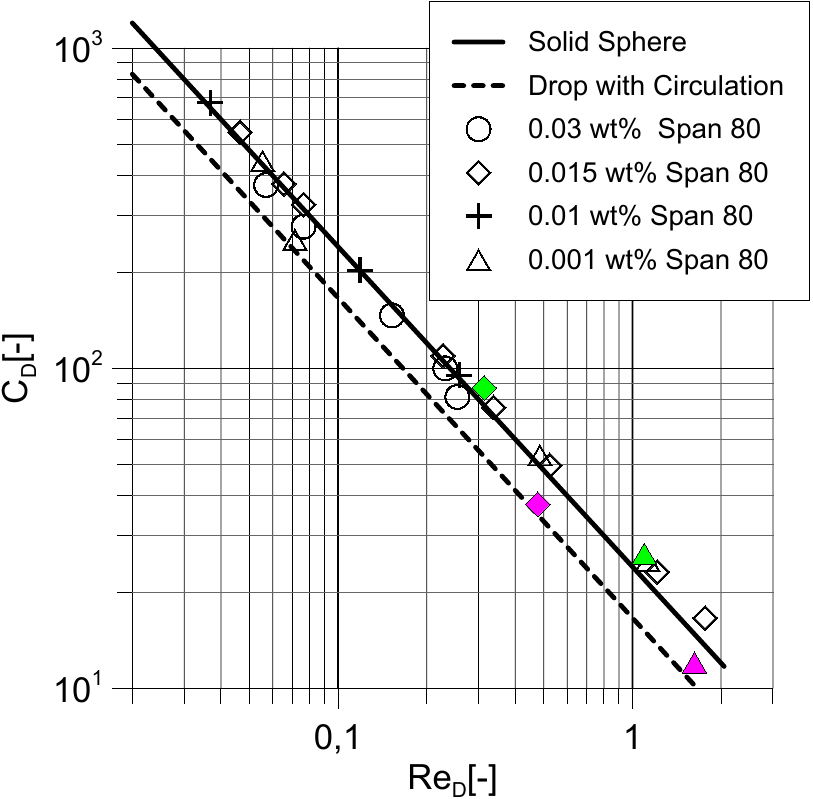}
  \caption{Plot of the drag coefficient versus Reynolds number. Unfilled symbols
  indicate experimental results. Filled green symbols indicate simulations with
  surfactant. Filled magenta symbols indicate equivalent simulations without
  surfactant.}
  \label{fig:Cd-vs-Re}
\end{figure}

As a consistency check, experimental results for the terminal velocity of
falling drops were compared with the theoretical values for drops with and without
a surface covering. This was done for several drop diameters and surfactant
concentrations, the results are shown in \cref{fig:Cd-vs-Re} (unfilled symbols).
As this figure shows, the results fall on or close to the line for a solid
sphere, which is as expected for a surfactant-covered drop. The dotted line
shows the theoretical result for a drop with circulation according to the
Hadamard-Rybczynski theory.

Some results from simulations of falling drops are also shown, using the same
shapes but with filled symbols. The points plotted in green are simulations with
surfactant-covered drops. These were done for a 1329 $\mu$m drop in 0.001 wt\%
Span 80 and for a 859 $\mu$m drop in 0.015 wt\% Span 80. The points plotted in
magenta are equivalent simulations without a surfactant present. It is seen that
simulations without surfactant agree well with the Hadamard-Rybczynski theory,
while simulations with surfactant agree well with the Stokes theory and with the
experimental results.

These simulations were done with a $20R$\x$40R$ computational domain and a grid
resolution of 400\x800. Grid refinement studies indicated that this was
sufficient. A moving grid procedure was employed, such that the drop was always
close to the center of the domain. The simulations were performed using the
``laboratory'' reference frame, where all velocities approach zero at the
boundary conditions, which were identical on all sides of the domain. That is to
say, full slip (zero gradient) was used for the tangential velocity while the
normal velocity (through the boundaries) was set to zero.

\subsection{Drop deformation}
Several experiments were performed with different drop diameters, field 
strengths and surfactant concentrations. The parameters for each
case are summarized in \cref{tab:cases}.
The drop shapes were extracted from the movies of the experiments using
the Spotlight software. The ratio between the major axis $a$ and the
minor axis $b$ could then be determined. Analysis of the various
uncertainties that affected the measurements $a$ and $b$ was done according to
the procedure described in \cite{zhao2009phd}, and the uncertainty in $a/b$ was
then computed assuming Gaussian error propagation. Herein, the
fluid properties were assumed to be without error. This gave a relative error 
in $a/b$ that was independent of $a/b$ but dependent on the initial radius. The
relative error was 2.0 \% for the largest drops considered here and 3.4 \% for 
the smallest.

 \begin{table}[!t]
  \caption{Cases studied for drop oscillation}
  \label{tab:cases}
  \centering
\begin{tabular}{cccc}
\toprule
  Case & Diameter [$\mu$m] & Span 80 [wt\%] & Field [V/mm] \\
\midrule
1 & 958 & 0.015 & 400.0 \\
2 & 826 & 0.015 & 533.3 \\
3 & 804 & 0.015 & 666.7 \\
4 & 560 & 0.015 & 266.7 \\
5 & 783 & 0.001 & 400.0 \\
\bottomrule
\end{tabular} 
\end{table}

\begin{figure}[!t] 
   \centering
   \includegraphics[width=0.93\linewidth]{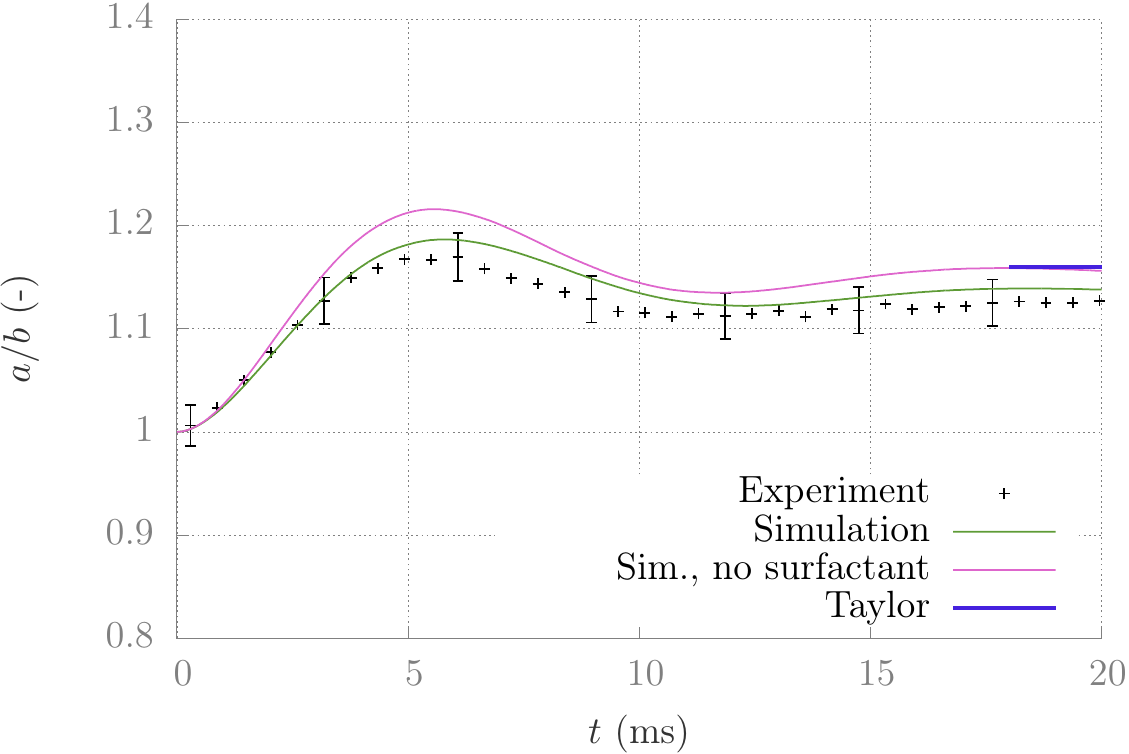}
   \caption{\textbf{Case 1:} Simulation and experimental results for the 958
     $\rm{\mu m}$ water drop falling through Marcol with 0.015 wt\% Span 80
   subjected to a 400.0 V/mm field}
   \label{fig:comp-958}
 \end{figure}

\begin{figure}[!t] 
   \centering
   \includegraphics[width=0.93\linewidth]{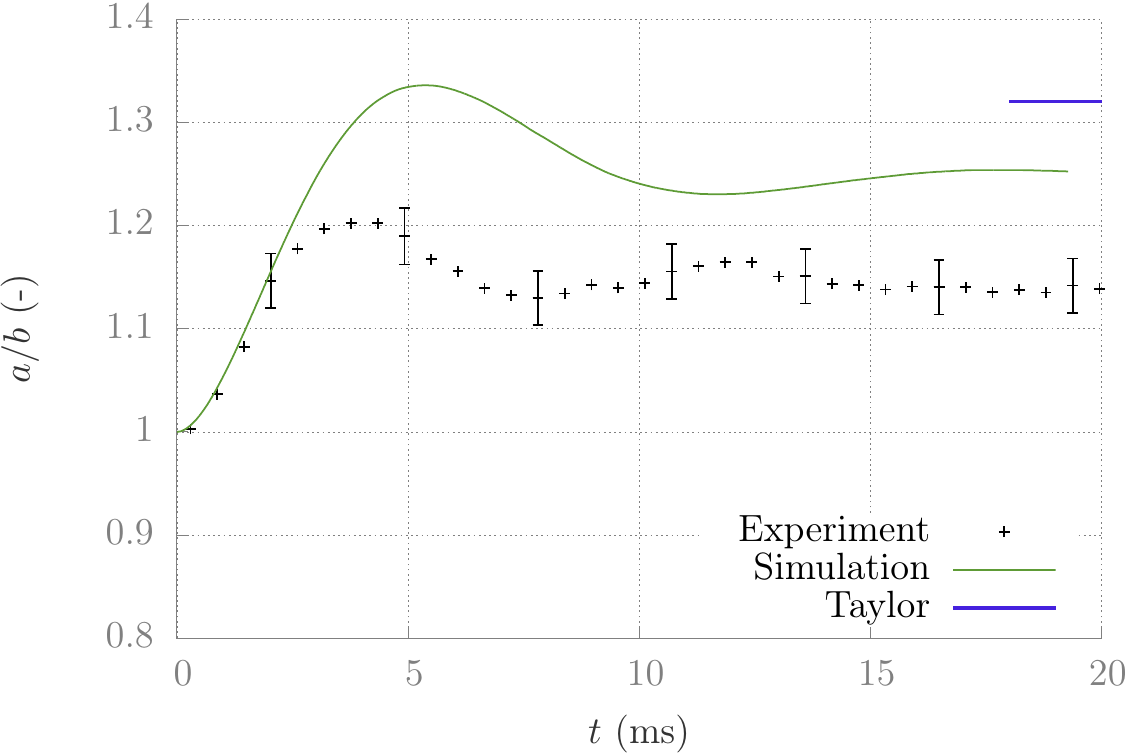}
   \caption{\textbf{Case 2:} Simulation and experimental results for the 826
     $\rm{\mu m}$ water drop falling through Marcol with 0.015 wt\% Span 80
   subjected to a 533.3 V/mm field}
   \label{fig:comp-826}
 \end{figure}

\begin{figure}[!t] 
   \centering
   \includegraphics[width=0.93\linewidth]{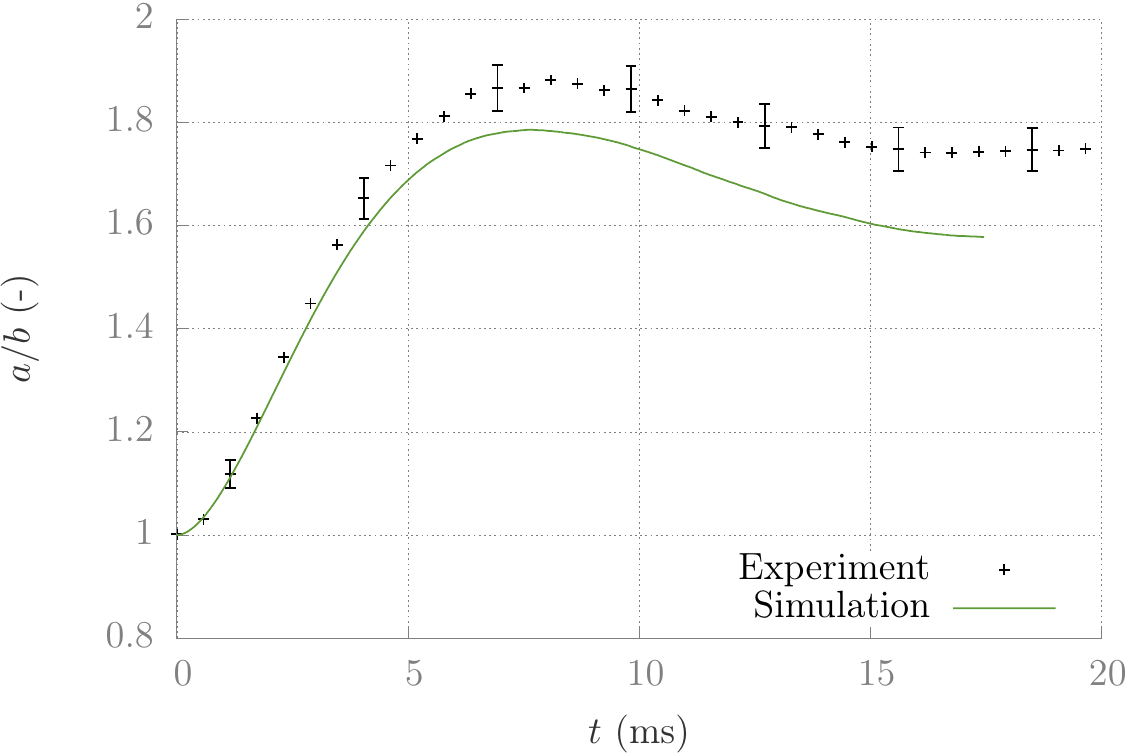}
   \caption{\textbf{Case 3:} \emph{Note: different scale!} Simulation and experimental results for the 804
     $\rm{\mu m}$ water drop falling through Marcol with 0.015 wt\% Span 80
   subjected to a 666.7 V/mm field}
   \label{fig:comp-804}
 \end{figure}

\begin{figure}[!t] 
   \centering
   \includegraphics[width=0.93\linewidth]{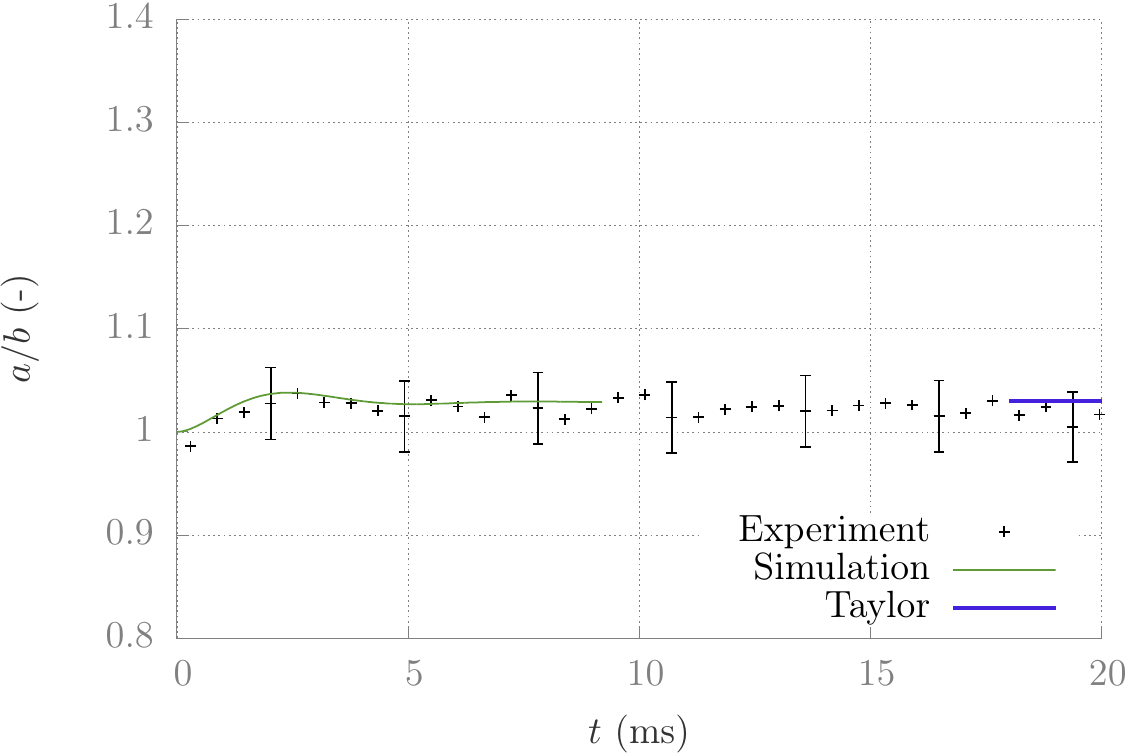}
   \caption{\textbf{Case 4:} Simulation and experimental results for the 560
     $\rm{\mu m}$ water drop falling through Marcol with 0.015 wt\% Span 80
   subjected to a 266.7 V/mm field}
   \label{fig:comp-560}
 \end{figure}

\begin{figure}[!t] 
   \centering
   \includegraphics[width=0.93\linewidth]{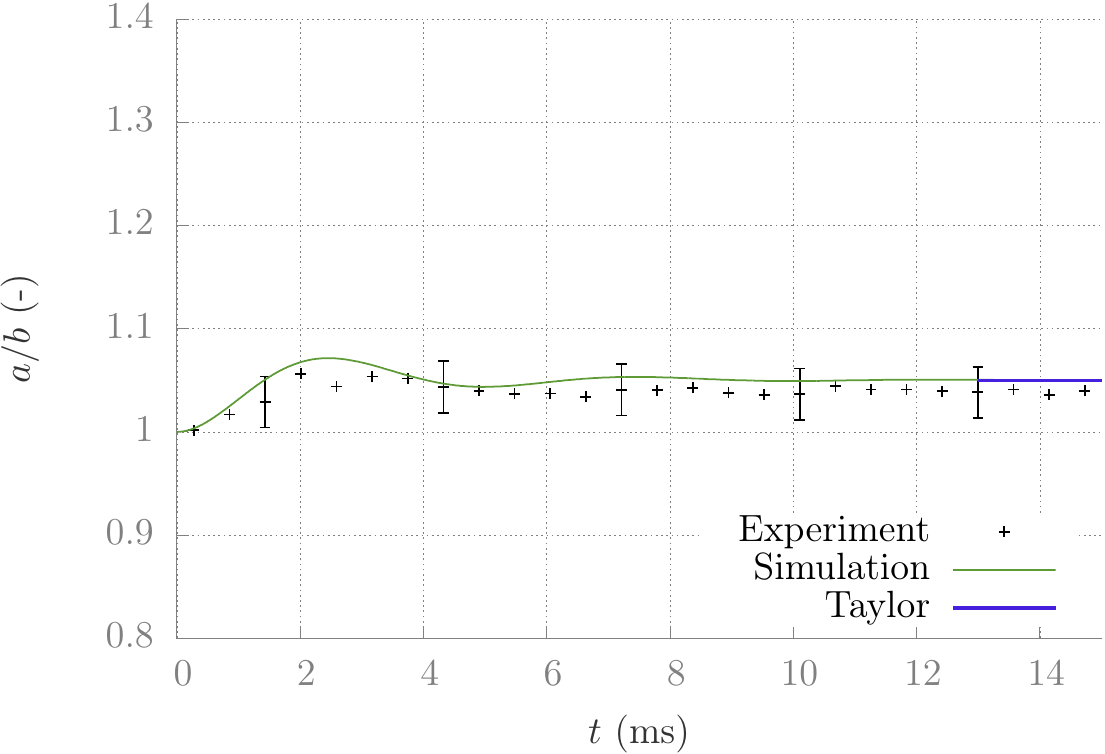}
   \caption{\textbf{Case 5:} Simulation and experimental results for the 783
     $\rm{\mu m}$ water drop falling through Marcol with 0.001 wt\% Span 80
   subjected to a 400.0 V/mm field}
   \label{fig:comp-783}
 \end{figure}

Based on these experiments, equivalent simulations were set up and run. The
values of $a/b$ were also extracted from the simulation results, such that
they could be plotted and compared directly to the experimental results. Such
plots are shown in in
\cref{fig:comp-958,fig:comp-826,fig:comp-804,fig:comp-560,fig:comp-783}. As is
seen in these plots, the agreement between simulations and experiments is good
for a variety of parameters. Error bars are shown for the experimental values
for every fifth point. The static deformation predicted by the Taylor theory
\cite{taylor1966} is shown with blue lines at the right-hand side. For Case
1 it is seen that this line agrees very well with the static deformation
predicted by simulations without surfactant, thus it overpredicts the
deformation in the presence of surfactant. 

Note that for Case 3 (\cref{fig:comp-804}), the deformation is much larger so
the scale is different.  Since the Taylor theory is only valid for small
deformations, it has not been applied here. For this case there is some
disagreement between experiments and simulations regarding the maximum
deformation and for the static deformation. This is likely due to the
assumption of insoluble surfactant in the simulation: as the drop is stretched
to such a large deformation, the interfacial area increases and there is room
for more surfactant at the interface. Thus adsorption kinetics can become
important in this case, making the simulations underpredict the deformation.

For Case 2 (\cref{fig:comp-826}) there is also a poor agreement. The cause of
this is harder to identify, since the other similar cases gave good agreement.
For this case, the difference between experiments and the Taylor theory is
also much larger than for the other cases. Further investigations are needed
to clarify this point.

For Case 1 (\cref{fig:comp-958}) a plot is also shown for a simulation without
surfactant, but using the experimentally measured value for the interfacial
tension at 0.015 wt\% Span 80 (magenta curve). It is seen that this simulation
overpredicts both the initial oscillation and the static deformation. This
indicates that the Marangoni forces due to the surfactant are important both
for the maximum and the static deformation.

In these simulations the grid resolution was 241\x482 in the radial and axial
directions, respectively. Simulations using coarser
grids indicated that this resolution was sufficient. All simulations were done
in axisymmetric coordinates, corresponding to a plane intersecting half the
drop. Initial simulations were performed with falling drops that had reached
terminal velocity before the field was applied. These were compared with 
simulations done with a field applied to a drop in zero gravity in a quiescent 
ambient fluid, and it was found that the difference in deformation was small.
Since the initial simulation with the falling drop reaching terminal velocity
has a runtime of many days, subsequent drop deformation simulations were done
for drops in zero gravity.

For these simulations without gravity, the full domain was 8$R$\x16$R$. The
boundary conditions used were identical on all sides of the domain, namely full
slip (zero gradient) for the tangential velocity while the normal velocity
(through the boundaries) was set to zero. The pressure boundary condition is
zero gradient in the normal direction, \ie $\partial p/\partial \v{n} = 0$.

\begin{figure}[!t] 
   \centering
   \includegraphics[width=\linewidth]{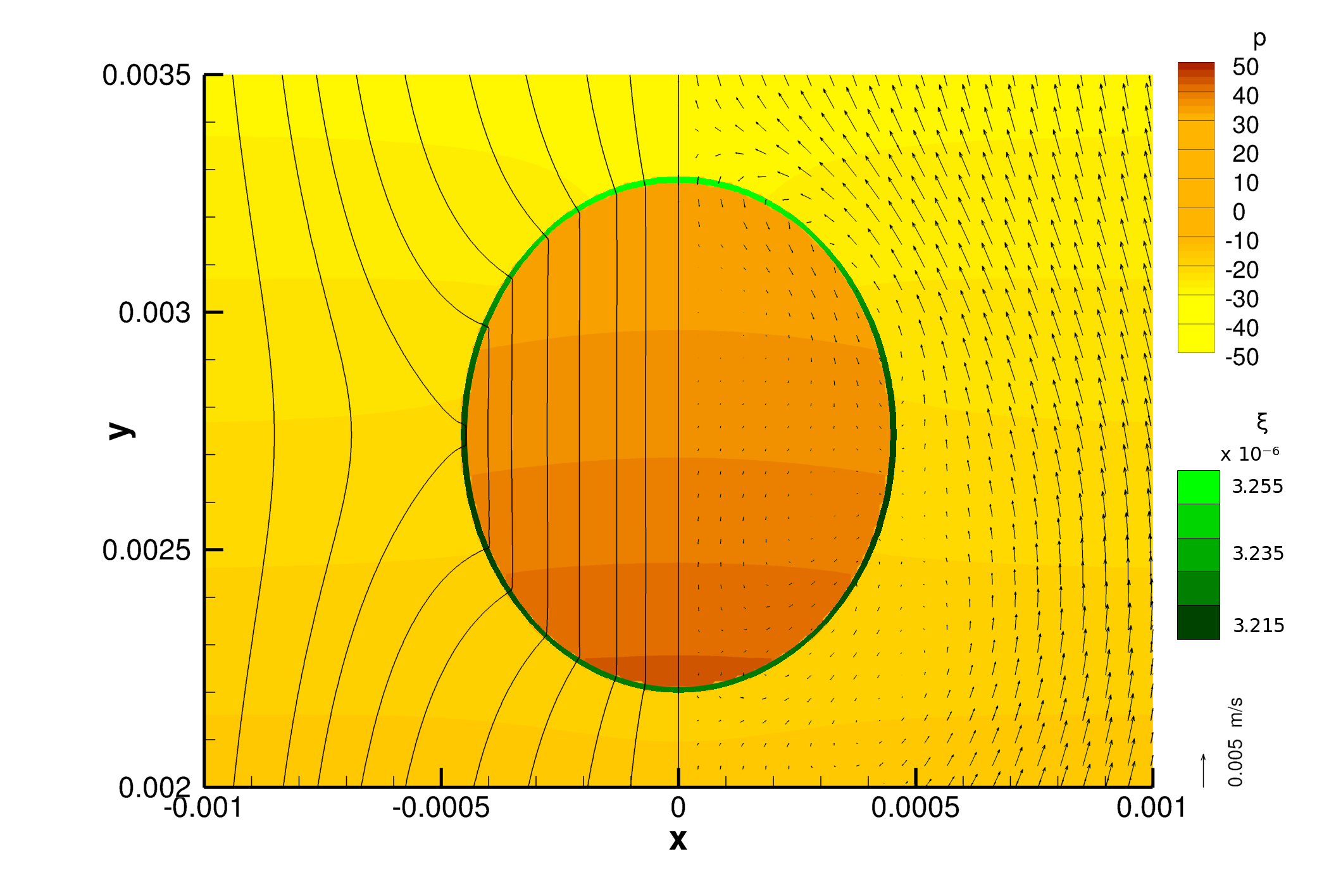}
   \caption{Detailed snapshot from the simulation of Case 1, the drop shown
   at its greatest extension. On the left-hand side, electric field lines are
   shown. The orange color shows the pressure field. The green colour on the 
   interface shows the surfactant concentration; the
   maximum concentration $\xi_\infty$ is 4.18$\cdot 10^{-6}$ for this case.
   \vspace{-5pt}}
   \label{fig:958-flow-efield}
 \end{figure}

A detailed view of the simulation result from Case 1 is shown in
\cref{fig:958-flow-efield}. The axisymmetric solution is shown mirrored about
the symmetry axis ($y$-axis). The drop is at its
most extended. The terminal velocity has been subtracted from the velocity
field, such that it appears in the drop reference
frame.  The orange colour indicates the pressure field and the
green color shows the surfactant distribution. The vectors
on the right-hand side show the velocity field,  and the field lines on the
left-hand side show the electric field. A reference vector for the velocity
is shown in the lower right corner. The simulation domain was
significantly larger than the view shown.

From this figure it is seen that the velocity field inside the droplet is
almost vanishing. This is reasonable, since the velocity field must be close
to zero as it reverses direction. It is also seen that the surfactant
distribution is slightly non-uniform, giving rise to Marangoni forces that
counteract the velocity field tangential to the interface. The electric field
lines shown on the left-hand side pass straight through the droplet, which has
high conductivity, and otherwise appear as one would expect.

 \vspace{-5pt}

 \section{Conclusions}
By experiments and computations we have investigated the
behaviour of single water drops in Marcol model oil with added
surfactant subjected to an electric field. This system is
well-characterized in terms of bulk, drop and interface properties. In
particular, the terminal velocity and the response to a sudden step in the
electric field have been studied.

In our computations, the details of the two-phase flow are captured using the
level-set method, and the interfacial jumps are accounted for using
the ghost-fluid method (GFM). The surfactant is modelled as insoluble and
the fluids are assumed to be perfect dielectrics.
The calculated terminal velocities for a clean system agree well with the
predictions of the Hadamard-Rybczynski formula. When the presence of
surfactant is accounted for, the calculated terminal velocities agree well
with the experimentally observed ones, which again agree with those
calculated using the Stokes formula for rigid bodies.

Our calculations of the dynamic drop stretching as a response to a
step change in the electric field agree well with the laboratory observations
(except for one case). Our results indicate that it is necessary to account
for the presence and dynamics of surfactant in order to reproduce the
dynamic behaviour of the drop.

\vspace{-5pt}

\section*{Acknowledgments} 
We would like to thank Dr. Martin Fossen (SINTEF Petroleum Research) for the
measurements of interfacial tension as a function of surfactant concentration,
Dr. Velaug Myrseth Oltedal (SINTEF Petroleum Research) for the measurements of
bulk viscosity, and Dr. Cédric Lesaint (SINTEF Energy Research) for the
measurements of density. We are also grateful to Dr. Pierre Atten and
Prof. Jean-Luc Reboud (G2ELab) and Dr. Erik Bjørklund (Wärtsilä) for fruitful
discussions around the work presented here.

This work was funded by the project \emph{Fundamental understanding of
electrocoalescence in heavy crude oils} coordinated by SINTEF Energy Research.
The authors acknowledge the support from the Petromaks programme of the
Research Council of Norway (206976), Petrobras, Statoil and W\"{a}rtsil\"{a}
Oil \& Gas Systems.

\vspace{-5pt}

\bibliographystyle{IEEEtran}
\bibliography{IEEEabrv,references}

\end{document}